LATEX file

\documentstyle[12pt]{article}

\def\lromn#1{\uppercase\expandafter{\romannumeral#1}}

\begin{document}

\begin{flushright}
TU/98/534\\
\end{flushright}

\vspace{12pt}

\begin{center}
\begin{large}
\renewcommand{\thefootnote}{\fnsymbol{footnote}}
\bf{
Prolonged Decoupling
}
\footnote[2]
{
Invited talk given at 3rd RESCU Symposium on "Particle Cosmology",
held at Tokyo,\\ November 11-13, 1997.
To appear in the Proceedings (Universal Academy Press, Tokyo).
}
\end{large}

\vspace{36pt}

\begin{large}
M. Yoshimura

Department of Physics, Tohoku University\\
Sendai 980-77 Japan\\
\end{large}

\vspace{4cm}

{\bf ABSTRACT}
\end{center}

We discuss decay of unstable particles and
pair annihilation of stable heavy particles that occur in the cosmic
medium, from the point of the fundamental microscopic theory. 
A fully quantum mechanical treatment shows that the effect
of thermal environment on these processes cannot be described in
terms of quantities on the mass shell alone, thus requiring an extention
of the Boltzmann-like equation.
The off-shell effect tends to prolong physical processes that take place
subsequent to the decay.

%%%%%%%%%%%%%%%%%%%%%%%%%%%%%%%%%%%%%%%%%%%%%%%%%%%%%

\newpage

\vspace{0.5cm} 
{\bf 1. Introduction}

\vspace{0.5cm} 
Any important event in cosmology is a series of elementary processes
that occur in the expanding universe.
Some of these events can be understood in terms of a rather simple 
process or their combination.
One should however keep in mind that the process does not occur
in isolation, unlike a well-prepared experiment in laboratory.
In the past the effect of environment on elementary processes
has been dealt with in very simplifed manners
in practical computation in cosmology, and 
in many cases has been completely ignored. 

We address this problem here and would like to fully incorporate the
quantum mechanical principles.
A subtle quantum mechanical effect might require a change of the
cherished practice in this field, 
but we would like to think of the problem, going back to the basic point.
It turns out that the familiar Boltzmann equation must be reexamined
due to its neglect of the off-shell effect.\cite{jmy-96}

Although no specific application is discussed in any detail here,
it is presumably helpful to mention at the outset
what I have in mind for applications.
They are X-boson decay for baryogenesis, 
and WIMP or LSP annihilation for dark matter.
In all these cases it is important that decay or
annihilation products make up a part of the cosmic environment.
Thus, when heavy particles 
are about to decay or decouple, inverse processes must
be considered simultaneously.
Interaction with thermal 
environment becomes the major concern for this problem.

A usual computational practice of the rate calculation for a particular 
process goes something like this;
one first computes the elementary S-matrix element,
\( \:
\langle f|S|i \rangle \,, 
\: \)
between the initial state $i$ and the final state $f$ on the mass shell, 
and then averages the probability in a thermal bath,
\( \:
\overline{|\langle f|S|i \rangle|^{2}} \,.
\: \)
This rate is used in the Boltzmann equation in the expanding universe.

I can think of two basic problems in this approach;
first, the states chosen $|i\rangle $ and $|f\rangle $ may not be
necessarily on the mass shell, as will be explained more fully below,
and secondly, even the on-shell matrix element
\begin{eqnarray*}
\langle f|T\,\exp \left( -i\,\int\,dt\,H_{{\rm int}}(t)\right)|i \rangle
\end{eqnarray*}
may contain propagation effects in thermal environment, different
from those in vacuum.

I shall first repeat the argument already made 
on the first point.\cite{weldon}
For this it is useful to look at the thermal Green function,
\begin{equation}
G_{{\rm th}}(\omega ) = \Delta (\omega + i\,\epsilon (\omega ))
- i\frac{\rho (|\omega |)}{e^{\beta |\omega |} - 1} \,, 
\end{equation}
with $T = 1/\beta $ the temperature and the spectral function is given by
\begin{equation}
\rho (\omega ) = \frac{2\,\Im \Pi (\omega )}{(\omega ^{2} - \omega _{k}^{2}
- \Re \Pi (\omega ))^{2} + (\,\Im \Pi(\omega ) \,)^{2}} \,,
\end{equation}
with $\omega _{k} = \sqrt{k^{2} + M^{2}}$.
Here $\Delta $ is the Green function in vacuum.
The proper self-energy $\Pi (\omega )$ is in general a complicated
object, and only in the weak coupling limit one may use the
Breit-Wigner function for $\rho (\omega )$, 
leading to the familiar form,
\begin{equation}
\rho_{{\rm BW}} (\omega ) = \frac{4\omega _{k}\,\Gamma }
{(\omega ^{2} - \omega _{k}^{2})^{2} + (2\omega _{k}\,\Gamma )^{2}}
\:\rightarrow  \:
G_{{\rm th}}(\omega )  \approx \Delta 
- \,\frac{2\pi i\,\delta (\omega ^{2} - \omega _{k}^{2})}
{e^{\beta \omega _{k}} - 1} \,.
\end{equation}
If one further assumes that the relevant $\omega $ integral
fully includes the peak region $\approx \omega _{k}$
of width $\Gamma $, then the Boltzmann equation follows.

The off-shell effect appears in two ways;
first, in the effective cutoff of the $\omega $ region,
\( \:
\omega > \omega _{c}
\: \)
(the threshold and taken to be $< \omega _{k}$) for $\omega $ integral.
If the temperature is below the threshold, $T < \omega _{c} $, 
the dominant region that contributes, $\omega < T$,
misses the peak region.
Second, if the coupling is not very weak, one must use the correct
form and not its Breit-Wigner approximation of the the spectral function
$\rho (\omega )$.
For instance, the behavior of $\Im \Pi (\omega )$ near the threshold
$\omega _{c}$ may become important at low temperatures.
A systematic treatment of the off-shell effect overcoming these points
is called for.

Let us make clear the fundamental difference between the on-shell
and the off-shell contributions.
In elementary quantum mechanics the off-shellness appears in
virtual states in perturbation formulas.
It deals with the short-time behavior and thus represents
any transition that does not conserve energy.
If one considers a decay of unstable system of mass $M$ and  decay
width $\Gamma $, then the off-shell effect is essential at times of
\( \:
t \approx 1/M
\: \)
while the on-shell S-matrix element well describes the 
long-time behavior of
\( \:
t \geq 1/\Gamma 
\: \)
(the lifetime).
As is well known, the on-shell behavior is given by the S-matrix,
while the off-shell effect must be discussed in terms of the more
fundamental quantity, the Green function.

%\vspace{1cm} 
\newpage
{\bf 2. Model of unstable particle decay}

\vspace{0.5cm} 
We consider here the case in which one can clearly separate
the small system in question and the cosmic environment.
This means that one should be able to ignore mutual interaction among 
unstable particles and also to ignore the change of the environment due to
the interaction with the unstable particle.
Thus, all the nonlinear effect and the back reaction is ignored.

Within this linear approximation one can think of a harmonic environment
of arbitrary spectral distribution
as well as a bilinear interaction between the small system and
the environment.
We then take the following model Hamiltonian,
\begin{eqnarray}
H &=& H_{c} + H_{b} + H_{{\rm int}} \nonumber 
\\ 
&=&
\omega _{k}\,c^{\dag }c 
+ \int_{\omega _{c}}^{\infty }\,d\omega \,\omega\, b^{\dag }(\omega )
b(\omega ) + \int_{\omega _{c}}^{\infty }
\,d\omega \,\sqrt{\sigma (\omega )}
\,\left( \,b^{\dag }(\omega )c + c^{\dag }b(\omega )\,\right) \,.
\end{eqnarray}
Here $c^{\dag } \,, c$ are creation and annihilation operators of
the unstable particle of energy $\omega _{k} = \sqrt{k^{2} + M^{2}}$,
and $b^{\dag }(\omega ) \,, b(\omega )$ are
those of the environment state, usually taken to be two-body states
of the decay product.
The threshold $\omega_{c} = \sqrt{k^{2} + (m_{1} + m_{2})^{2}}$ and
$m_{i}$ is the masses of decay product.
The instability condition is imposed; $\omega _{c} < \omega _{k}$.

Although for definiteness we consider the unstable particle decay, one
may choose with some generalization a pair of heavy stable particles
for an initial state.

It should be reminded of that even in this linear model the irreversibility
exists; it can occur because an unstable particle
couples to an infinitely many degenerate environment states.
A great virtue of this model is that it can be analytically solved
\cite{jmy-97-1}
and in principle one has many explicit results that can test 
approximation schemes one always needs in practice.
We shall discuss extention of this model later.

A way to make integrability explicit is to present the operator
solution.
One can find variables that diagonalize the Hamiltonian;
\begin{eqnarray}
&& \hspace*{-1cm}
B^{\dag }(\omega ) = b^{\dag }(\omega ) +
F(\omega + i0^{+})\,\left( \,-\,\sqrt{\sigma (\omega )}\,c^{\dag }
+ \int_{\omega _{c}}^{\infty }\,d\omega \,
\frac{\sqrt{\sigma (\omega )\sigma (\omega ')}}{\omega ' - \omega - i0^{+}}
\,b^{\dag }(\omega ')\,\right) \,,
\end{eqnarray}
where 
\begin{eqnarray*}
H = \int_{\omega _{c}}^{\infty }\,d\omega \,\omega \,B^{\dag }(\omega )
B(\omega ) \,.
\end{eqnarray*}
The transformation from $c\,, b(\omega )$ to $B(\omega )$ 
(and their conjugates) is canonical. 
Even the operator inversion can be written down.
The important function $F(z)$ is
analytic except along the cut, $z > \omega _{c}$, and explicitly
\begin{eqnarray}
F(z) = \frac{1}{-\,z + \omega _{k} - \int_{\omega _{c}}^{\infty }\,
d\omega \,\frac{\sigma (\omega )}{\omega - z}} \,.
\end{eqnarray}
We do not explain here the frequency and the wave function
renormalization, for which we refer to our paper.\cite{jmy-97-1}

The essential reason why this model is integrable is saturation of
unitarity by the elastic one. 
Analogy to the scattering problem is useful and is indeed made
very evident by considering the energy eigenstate,
\begin{eqnarray*}
|\omega \rangle _{S} = B^{\dag }(\omega )|0\rangle \,, 
\end{eqnarray*}
where $|0\rangle $ is the perturbative (and also the true) vacuum with
\( \:
c|0\rangle = 0 \,.
\: \)
Clearly, this state $|\omega \rangle _{S}$ is a linear superposition of
\( \:
c^{\dag }|0\rangle 
\: \)
and 
\( \:
b^{\dag }(\omega )|0\rangle \,.
\: \)
When $\omega _{k} > \omega _{c}$, 
the original one-particle state $c^{\dag }|0\rangle$ 
becomes a resonance, or the unstable particle, 
which is $|\omega \rangle_{S}$.
The overlap probability
\begin{eqnarray*}
|\langle 0|c|\omega  \rangle_{S}|^{2} \equiv H(\omega )
\end{eqnarray*}
is very fundamental.
It is related to the discontinuity along the cut,
\begin{eqnarray}
&&
F(\omega + i0^{+}) - F(\omega - i0^{+}) = 2\pi i\,
\sigma (\omega )|F(\omega + i0^{+})|^{2} = 2\pi i\,H(\omega ) \,.
\end{eqnarray}
In terms of the spectral function
\begin{equation}
H(\omega ) = \frac{\sigma (\omega )}{(\omega - \omega _{k}
+ \Re \Pi (\omega ))^{2}
+ (\pi \,\sigma (\omega))^{2}} \,,
\end{equation}
where 
\begin{eqnarray*}
\Pi (z) = \int_{\omega _{c}}^{\infty }\,d\omega \,\frac{\sigma (\omega )}
{\omega - z} \,.
\end{eqnarray*}
From the discontinuity structure it follows that a simple pole
exists in the second Riemann sheet extended via this discontinuity formula;
the original state pole on the real axis at $\omega = \omega _{k}$
moves into the second sheet and acquires an imaginary part,
\( \:
\approx \Gamma = 2\pi \sigma (E_{r})
\: \)
where $E_{r}$ is the real part of the pole (in the weak coupling limit,
$E_{r} \approx \omega _{k}$ after renormalization).
This pole describes the exponential decay law; the non-decay amplitude
\( \:
\approx e^{-i\omega _{k}t - \Gamma t/2} \,.
\: \)

It can also be shown \cite{jmy-97-1}
that this quantity $H(\omega )$ is directly related
to the Lehmann spectral function $\rho_{L} (\omega )$ by
\begin{eqnarray*}
H(\omega ) = \frac{(\omega + M)^{2}}{2M}\,\rho_{L} (\omega ) \,.
\end{eqnarray*}

The decay law of the unstable particle state 
$|1\rangle = c^{\dag }|0\rangle $
in vacuum is precisely described as follows.
One notes that the non-decay amplitude,
\begin{eqnarray*}
\langle 1|e^{-iHt}|1 \rangle = \langle 0|c\,e^{-iHt}\,c^{\dag }|0 \rangle
\,, 
\end{eqnarray*}
is calculable if one finds the Heisenberg evolution,
\( \:
c(t) = e^{iHt}\,c\,e^{-iHt} \,.
\: \)
Since the operator relation is known, this is given by
\begin{equation}
c(t) = -\,\int_{\omega _{c}}^{\infty }\,d\omega \,\sqrt{\sigma (\omega )}
\,F(\omega + i0^{+})e^{-i\omega t}\,B(\omega ) \,.
\end{equation}
It is useful to rewrite the Heisenberg operator in terms of the original
variables;
\begin{eqnarray}
&&
c^{\dag }(t) = g(t)\,c^{\dag } + i\,
\int_{\omega _{c}}^{\infty }\,d\omega \,
\sqrt{\sigma (\omega )}\,h(\omega \,, t)e^{i\omega t}\,b^{\dag }(\omega )
\,.
\label{heisenberg operator} 
\end{eqnarray}
The basic function here is $g(t)$;
\begin{eqnarray}
&&
g(t) = \int_{\omega _{c}}^{\infty }\,d\omega \,H(\omega )e^{i\omega t}
= \frac{1}{2\pi i}\,\int_{C_{0} + C_{1}}\,dz\,F(z)\,e^{izt} 
\,, 
\\ &&
h(\omega \,, t) = i\,F^{*}(\omega + i0^{+}) - ik(\omega \,, t) \,, 
\hspace{0.5cm} 
\dot{k}(\omega \,, t) = i\,e^{-i\omega t}\,g(t) \,.
\end{eqnarray}
The boundary condition for $k(\omega \,, t)$ is
\( \:
k(\omega \,, t) \rightarrow 0
\: \)
as $t \rightarrow \infty $, hence
\begin{eqnarray}
&&
k(\omega \,, t) =
\frac{1}{2\pi i}\,\int_{C_{0} + C_{1}}
\,dz\,\frac{e^{i(z - \omega )t}}{z - \omega }\,F(z) \,.
\end{eqnarray}
The non-decay amplitude is given by
\begin{eqnarray}
\langle 1|e^{-iHt}|1 \rangle = g^{*}(t) =
\int_{\omega _{c}}^{\infty }\,d\omega\,\sigma (\omega )\,
|F(\omega + i0^{+})|^{2}\,e^{-i\omega t} \,.
\end{eqnarray}

The second form of the $\omega $ integral for $g(t)$ along $C_{0} + C_{1}$
makes separation of the pole term explicit; 
the contour $C_{0}$ encircles the pole (conjugate to the one already
discussed, due to $e^{izt}$ factor instead of $e^{-izt}$) 
in the second sheet, while
$C_{1}$ consists of two lines parallel to the imaginary axis, one
in the first and the other in the second sheet.
It is clear that the $C_{1}$ integral gives the non-exponential decay
law which is important both at early and late times of the decay.
For instance, at very late times the important part of $\omega $ integral
comes from the threshold region, and if one parametrizes this region
using the rate $\Gamma $,
\begin{equation}
\sigma (\omega ) \approx \frac{\Gamma }{2\pi }\,
\left( \frac{\omega - \omega _{c}}{Q}\right)^{\alpha } \,, 
\label{threshold spectrum} 
\end{equation}
the non-decay amplitude is given by
\begin{eqnarray}
&&
\langle 1|e^{-iHt}|1 \rangle  \approx -\,i\frac{\Gamma }{2\pi Q}\,
\frac{\Gamma (\alpha + 1)}{(Qt)^{\alpha + 1}}\,e^{-i\omega _{c}t
- i\pi \alpha /2}
\,,
\end{eqnarray}
with $Q = \omega _{k} - \omega _{c}$.
The decay at late times thus follows the power law;
\begin{eqnarray*}
\approx (\frac{\Gamma }{2\pi Q})^{2}(Qt)^{-2(\alpha + 1)} \,.
\end{eqnarray*}

\vspace{1cm} 
{\bf 3. Decay in cosmic medium}

\vspace{0.5cm} 
Amuzingly, this model and the basic technique outlined is taken over
to derive the decay law in thermal medium.\cite{jmy-96},
\cite{off-shell remnant and b-asymmetry}
In the thermal medium one is interested in time evolution of the occupation
number, $f(t)$, to be given by the thermal average over a given
mixed state,
\begin{equation}
f(t) \equiv 
\langle c^{\dag }(t)c(t) \rangle_{i} = 
{\rm tr}\;(\,c^{\dag }(t)c(t)\,\rho _{i}\,) \,, 
\end{equation}
where $\rho _{i}$ is the density matrix of the mixed state.

We take for $\rho _{i}$ the thermal density matrix,
\begin{eqnarray*}
\rho _{i} = e^{-\,\beta (H_{b} + H_{c})}/({\rm tr}\;
e^{-\,\beta (H_{b} + H_{c})}) \,,
\end{eqnarray*}
where $H_{b}$ and $H_{c}$ are the environment and the subsystem
parts of the Hamiltonian.
It is perhaps necessary to explain why we take for the initial mixed state
this one and not the
one like $e^{-\beta H}$ using the total Hamiltonian $H$.
The cosmic environment we consider is not a stationary state; it
adiabatically changes with the cosmic expansion.
With the momentum redshift a time lag occurs during the decay
between the unstable particle and its lighter decay products.
Light decay products have their own stronger interaction not written in 
the Hamiltonian above, and they tend to keep in thermal equilibrium
with the well-known temperature variation,
\( \:
T(t)a(t) =
\: \)
constant, with $a(t)$ the cosmic scale factor.
On the other hand, the decay and its inverse process cannot keep pace
with the cosmic expansion.
We model this circumstance by postulating that the environment by itself
is always in thermal equilibrium, 
while the unstable particle starts to lose
its thermal correlation with the environment.
This is the reason we take the initial density matrix uncorrelated
between the unstable and the decay product particles, although
at much higher temperatures they may be in thermal contact.
Thus, we take for the initial state average
\begin{eqnarray}
&&
\langle c^{\dag }c \rangle = f(0) \,, \hspace{0.5cm} 
\langle c^{\dag }b(\omega ) \rangle_{i} = 0 \,,
\hspace{0.5cm} 
\langle b^{\dag }(\omega )b(\omega ') \rangle_{i}
= \delta (\omega - \omega ')\,\frac{1}{e^{\beta \omega } - 1} \,.
\end{eqnarray}

For the moment we ignore the effect of the cosmic expansion.
Time evolution of the occupation number $f(t)$ can be derived 
using the explicit form of $c^{\dag }(t)$, eq.(\ref{heisenberg operator});
\begin{equation}
f(t) = |g(t)|^{2}\,f(0) + \int_{\omega _{c}}^{\infty }\,d\omega \,
\frac{\sigma (\omega )|h(\omega \,,t)|^{2}}{e^{\beta \omega } - 1} \,.
\end{equation}
Its time derivative can be written in the following form,
\begin{equation}
\frac{df}{dt} = -\,\Gamma (t)\,|g(t)|^{2}\,f(0) + 
\int_{\omega _{c}}^{\infty }\,d\omega \,\sigma (\omega )\,
\frac{2\Re (gh^{*}e^{-i\omega t})}{e^{\beta \omega } - 1} \,, 
\label{time evolution eq1} 
\end{equation}
where the time dependent decay rate is defined by
\begin{equation}
\Gamma (t) = - 2\Re \frac{\dot{g}}{g} = -\,\frac{d}{dt}\ln |g|^{2} \,.
\end{equation}
The rate $\Gamma (t)$ is time independent during the pole dominant epoch,
but at late times it decreases as
\( \:
\Gamma (t) \rightarrow \frac{2(\alpha + 1)}{t} \,,
\: \)
using the spectral function having the 
threshold behavior (\ref{threshold spectrum}).
An equivalent form for the time evolution equation
is obtained by eliminating the initial value ($f(0)$)
dependence;
\begin{equation}
\frac{df}{dt} + \Gamma(t) f =
\int_{\omega _{c}}^{\infty }\,d\omega \,\sigma (\omega )\,
\frac{2\Re (gh^{*}e^{-i\omega t}) + \Gamma (t)\,|h|^{2}}
{e^{\beta \omega } - 1}
\,.
\label{time evolution eq2} 
\end{equation}
This form is more convenient to application to cosmology.

The first form of time evolution equation (\ref{time evolution eq1})
has a clear interpretation
if one retains only the on-shell contribution; by using the pole
dominance for 
\( \:
g(t) = e^{i\omega _{k}t - \Gamma t/2} \,, 
\: \)
the equation is reduced to
\begin{equation}
\frac{df}{dt} = -\,\Gamma e^{-\Gamma t}\,\left( f(0)
- \frac{1}{e^{\beta \omega _{k}} - 1}\right) \,.
\end{equation}
There are oscillatory terms such as $e^{-\Gamma t/2} \times $ 
\( \:
\left( \,\cos \;{\rm or}\; \sin (\omega - \omega _{k})t\,\right) \,, 
\: \)
which averages out in the $\omega $ integral, leading to this result.
Thus, if the initial occupation $f(0)$
is chosen to be that of the thermal one,
then there is no change, while its deviation from the equilibrium value
is relaxed back by the rate $\Gamma $.

The second form of evolution equation (\ref{time evolution eq2})
suggests that irrespcetive of
the pole dominance the occupation number has an asymptote;
\begin{equation}
f_{\infty }  = \int_{\omega _{c}}^{\infty }\,d\omega \,
\frac{H(\omega )}{e^{\beta \omega } - 1} 
\approx \int_{k}^{\infty }\,d\omega \,
\frac{\sigma (\omega )}{(\omega - \omega _{k})^{2} + 
(\pi \sigma (\omega ))^{2}}\,\frac{1}{e^{\beta \omega } - 1}
\,, \label{asymptotic occ-n} 
\end{equation}
since both of $g(t)$ and $k(t)$ in $h(\omega \,, t)$ asymptotically vanish 
and
\begin{eqnarray*}
h(\omega \,, t) \rightarrow iF^{*}(\omega + i0^{+}) \,.
\end{eqnarray*}
This asymptotic form reveals the essence of our result;
only when the full Breit-Wigner region for $H(\omega )$ contributes,
the on-shell result gives a good description for the asymptotic
occupation number. Otherwise,
\( \:
f_{\infty } \neq \frac{1}{e^{\beta \omega _{k}} - 1} \,
\: \)
(on-shell result).

The particle number density is calculated by summing the occupation number
over independent momenta;
\begin{equation}
n(t) = \frac{1}{2\pi ^{2}}\,\int_{0}^{\infty }\,dk\,k^{2}\,f(t \,, k)
\,.
\end{equation}
Let us work out its asymptotic expression $n_{\infty }$, 
using $f_{\infty }$ for $f(t\,, k)$.
When the pole term dominates, this gives a well-known result;
for the non-relativistic region,
\begin{eqnarray*}
n_{\infty } \approx (\frac{MT}{2\pi})^{3/2}\,e^{-M/T} \,.
\end{eqnarray*}
On the other hand, at low temperatures the threshold region dominates
in the $\omega $ integral, and one has a power of temperature;
\begin{equation}
n_{\infty } \approx 
A(\alpha )\,\frac{\Gamma }{M}\,(\frac{T}{M})^{\alpha  + 1}\,T^{3} \,,
\hspace{0.5cm} 
A(\alpha ) = 
\frac{\zeta (\alpha + 4)\Gamma (\alpha + 4)\Gamma (\frac{\alpha }{2} + 1)}
{16\pi ^{2}\,\sqrt{\pi }\,\Gamma (\frac{\alpha }{2} + \frac{5}{2})} \,.
\label{off-sh n-density} 
\end{equation}
Thus, the low temperature behavior is much enhanced compared to that
given by the on-shell contribution alone.

The effect of the cosmic expansion is readily incorpolated for the evolution
of the number density. The easiest way is to write it for the dimensionless
yield $Y \equiv \frac{n}{T^{3}}$, using the dimensionless time variable,
\( \:
\tau \equiv \Gamma t \, ;
\: \)
\begin{eqnarray}
&&
\hspace*{-1cm}
\frac{dY}{d\tau } = 
-\,\int_{0}^{\infty }\,\frac{dk\,k^{2}}{2\pi ^{2}\,T^{3}}\,
\gamma (t\,, k)\,\left( \,f_{k}(t) - 
\,\int_{k}^{\infty }\,d\omega \,
\frac{\sigma (\omega \,, k)}{(\omega - \omega _{k} )^{2} 
+ (\Gamma /2)^{2}}\,\frac{1}{e^{\beta \omega } - 1} \,\right)\,,   
\\ &&
Y = \frac{1}{T^{3}}\,\int_{0}^{\infty }\,\frac{dk\,k^{2}}{2\pi ^{2}}\,
f_{k}(t) \,, 
\\ &&
\gamma (t\,, k) = -\,2\,\Re \,
\frac{d}{\Gamma \,dt}\ln (g_{0}(t \,, k) + g_{1}(t\,, k))
= -\,\frac{\frac{d}{\Gamma \,dt}|g_{0}(t \,, k) + g_{1}(t\,, k)|^{2}}
{|g_{0}(t \,, k) + g_{1}(t\,, k)|^{2}} \,.
\end{eqnarray}
The momentum dependent spectral is
\( \:
\sigma (\omega \,, \vec{k}) = \frac{M}{\omega _{k}}\,
\sigma (\sqrt{\omega ^{2} - \vec{k}^{2}})
\,,
\: \)
where $\sigma (\omega )$ is the one in the rest frame.
The usual time-temperature relation $T\propto 1/\sqrt{t}$ should
be used in this equation.

The asymptotic solution for this equation is readily found,
ignoring the time dilatation effect ($M/\omega _{k}\rightarrow 1$);
\begin{eqnarray}
&&
Y \:\rightarrow  \: 
\frac{(\alpha + 2)\zeta (\alpha + 4)\Gamma (\alpha + 3)
\Gamma (\frac{\alpha }{2} + 1)}
{8\pi ^{2}\,\sqrt{\pi }\,\Gamma (\frac{\alpha }{2} + \frac{5}{2})}
\,\frac{\Gamma }{M}\,(\frac{T}{M})^{\alpha + 1} \,.
\end{eqnarray}
The important point is that it has a power behavior of the cosmic
temperature, as is guessed from $n_{\infty }$ of 
eq.(\ref{off-sh n-density}).
The power $\alpha + 1$ is related to the threshold behavior of the
spectral function $\sigma (\omega )$.

I refer to our recent paper \cite{off-shell remnant and b-asymmetry} 
on the detailed time evolution, but
roughly $Y$ follows initially the thermal behavior,
\( \:
(\frac{M}{2\pi T})^{3/2}\,e^{-M/T} \,,
\: \)
and after a short transient epoch it approaches the asymptotic form above.
The transition epoch is characterized by two temperature scales,
$T_{*}$ and $T_{{\rm eq}}$ given by
\begin{eqnarray}
&&
\frac{T_{*}}{M} \approx \sqrt{\frac{\eta }{(\alpha + 4)\,
\ln \frac{M}{\Gamma }}} \,, \hspace{0.5cm} 
\eta = \sqrt{\frac{45}{16\pi ^{3}\,N}}\,\frac{m_{{\rm pl}}\Gamma }{M^{2}}
\,,
\\ &&
\frac{T_{{\rm eq}}}{M} \approx 
\left( \ln \frac{M}{\Gamma }\right)^{-1}
\,,
\end{eqnarray}
where $N$ is the number of massless particle species contributing to
the energy density.
The quantity $\eta $ is the decay rate relative the Hubble rate
calculated at $T = M$.
Usually, $T_{*} > T_{{\rm eq}}$.
The higher temperature $T_{*}$ is given by the time when the exponential
pole contribution is changed to the off-shell power contribution, while
$T_{{\rm eq}}$ is the one when the two temperature dependent terms 
in the stationary number density $n_{\infty }$ compete equally;
\begin{eqnarray}
&&
(\frac{M}{2\pi T_{{\rm eq}}})^{3/2}\,e^{-M/T_{{\rm eq}}} \approx 
\frac{\Gamma }{M}(\frac{T_{{\rm eq}}}{M})^{\alpha + 1} \,.
\end{eqnarray}

Importance of the off-shell effect is determined by the remnant density
at $T = T_{{\rm eq}}$;
\( \:
(n/T^{3})_{T = T_{{\rm eq}}} \,.
\: \)
Besides a weak logarithmic dependence, this quantity has the factor,
$\Gamma /M$. The effect is thus more pronounced for a larger rate
$\Gamma $, or a stronger coupling.

\vspace{1cm} 
{\bf 4. Some application and extention}

\vspace{0.5cm} 
Physical processes that occur after this transient time are prolonged,
compared to the instantaneous decay in the exponential law.
One of the important effects of this sort is baryogenesis at the GUT epoch.
In the scenario of a heavy $X$ boson decay the out-of-equilibrium 
condition was considered to impose a rather severe constraint on
particle physics and cosmology.\cite{b-asymmetry review}
This constraint arises due to that the abundance of $X$ and $\bar{X}$ bosons
should not be too much suppressed when the asymmetry generation starts.
The on-shell kinematics requires that the temperature $T_{d}$ at
the Hubble $H = \Gamma $ must not be too much larger than the mass of
$X$ boson; thus $T_{d} < m_{X}$. Otherwise, the inverse decay can
take place frequently, to suppress the number density of $X$ boson.
This leads to the lower bound of the $X$ boson mass,\\
\( \:
m_{X} > O[1]\times \frac{\alpha _{X}\,m_{{\rm pl}}}{\sqrt{N}}
\approx 10^{16}\,{\rm GeV} \,.
\: \)

This constraint is however based on the on-shell kinematics;
thermal quarks and leptons are energetically
difficult to produce heavy $X$ bosons
when the cosmic temperature $T < m_{X}$.
But, as we discussed here, heavy $X$ bosons that were once abundant
do not disappear immediately due to the off-shell effect.
Thus, one can anticipate that the above lower bound of the $X$ mass
is considerably relaxed.\cite{jmy-96} 
For details of this problem, 
I refer to our forthcoming paper.\cite{off-shell remnant and b-asymmetry}

As an extention of the model so far discussed, it is instructive
to treat the case of pair annihilation model.
The model can describe two particle annihilation into lighter particles,
\begin{eqnarray*}
c_{1} \,c_{2} \rightarrow b(\omega ) \,, 
\end{eqnarray*}
again $b^{\dag }(\omega )|0\rangle $ taken to be two-body states of
light particles. Its model Hamiltonian is
\begin{equation}
H = \omega _{1}\,c_{1}^{\dag }c_{1} + \omega _{2}\,c_{2}^{\dag }c_{2}
+ \int_{\omega _{c}}^{\infty }\,d\omega \,\omega \,b^{\dag }(\omega )
b(\omega ) + \int_{\omega _{c}}^{\infty }
\,d\omega \,\sqrt{\sigma (\omega )}\,\left( \,
b^{\dag }(\omega )c_{1}c_{2} + ({\rm h.c.}) \,\right) \,.
\end{equation}
This time the model does not allow exact solution, but one can use
the method of Feynman and Vernon \cite{feynman-vernon}, 
to integrate the effect of environment variables.
We leave details of this technical part to our future paper,
but let me discuss how one may derive a Boltzmann-like equation.
The key for this is to neglect a long-time correlation and to take the
$t \rightarrow \infty $ limit in order to erase the off-shell effect.
After cutting off the long-time correlation, we find that
\begin{eqnarray}
\frac{d}{dt}\langle c_{i}^{\dag }(t)c_{i}(t) \rangle &=&
\int_{\omega _{c}}^{\infty }\,d\omega \,
\sigma (\omega )\,\int_{0}^{t}\,ds\,
\nonumber \\ && \hspace*{-2cm}
2\,\cos (\omega - \omega _{1} - \omega _{2})s
\,\left( \,-\,f_{1}f_{2}(f_{{\rm th}} + 1) 
+ (f_{1} + 1)(f_{2} + 1)f_{{\rm th}} \,)\,\right)
\,,
\end{eqnarray}
where 
\( \:
f_{{\rm th}}(\omega ) = 1/(e^{\beta \omega } - 1)
\: \)
and $f_{i}(\omega _{i})$ are occupation numbers of $i$.
One may then use
\begin{equation}
\lim_{t \rightarrow \infty }\,\int_{0}^{t}\,ds\,
2\cos (\omega - \omega _{1} - \omega _{2})s = 2\pi \,\delta 
(\omega - \omega _{1} - \omega _{2}) \,, 
\end{equation}
to get
\begin{equation}
\frac{d}{dt}\langle c_{i}^{\dag }(t)c_{i}(t) \rangle =
\Gamma \left( \,-\,
f_{1}(\omega _{1})f_{2}(\omega _{2})(f_{{\rm th}} + 1)
+ (f_{1}(\omega _{1}) + 1)(f_{2}(\omega _{2}) + 1)f_{{\rm th}} \,\right)
\,,
\end{equation}
with
\( \:
f_{{\rm th}} = f_{{\rm th}}(\omega = \omega _{1} + \omega _{2}) \,.
\: \)
This is the Boltzmann equation with the rate averaged in thermal medium.
Our formalism can thus describe the kinetic equation
for the pair annihilation in thermal medium, 
once the thermal distribution of 
two-body states $f_{{\rm th}}(\omega )$ is given.

\end{document}